# Some Relations between Divergence Derivatives and Estimation in Gaussian channels

Jacob Binia, Member, IEEE

*Abstract* — The minimum mean-square error of the estimation of a non-Gaussian signal where observed from an additive white Gaussian noise (WGN) channel's output, is analyzed. First, a quite general time-continuous channel model is assumed for which the behavior of the non-Gaussianess of the channel's output for small signal to noise ratio q, is proved. Then, It is assumed that the channel input's signal is composed of a (normalized) sum of N narrowband, mutually independent waves. It is shown that if N goes to infinity, then for any fixed q (no mater how big) both CMMSE and MMSE converge to the signal energy at a rate which is proportional to 1/N. Finally, a known result for the MMSE in the one-dimensional case, for small q, is used to show that all the first four terms in the Taylor expansion of the non-Gaussianess of the channel's output equal to zero.

## I. Introduction

In section II we show, under quite general time-continuous channel model, that the derivative at zero snr of the non-Gaussianess of the output equals zero.

Then (section III) it is proved that a signal that is composed of the sum of N independent CW waves, is hard to be estimated where observed from its sum with Gaussian noise, for large N. In this case both MMSE (Minimum Mean-Square Error) and CMMSE (Causal Minimum Mean-Square Error) of the appropriate optimal estimations tend to the signal energy in a rate that is proportional to 1/N. This is true for any snr, no matter how big but fixed.

In section IV we move in opposite direction. Based on a known result for the Taylor expansion of MMSE of a random variable X for small snr, it is shown that the values of the first three derivatives at zero snr, of the non-Gaussianess of the sum of a random variable X and a normal Gaussian variable W, equal to zero. An expression for the fourth derivative as a function of the third and the fourth moments of X is also given.

The author is with the New Elective–Engineering Services Ltd, 34970 Haifa, Israel (e-mail: biniaja@netvision.net.il).

## II. Non-Gaussianess of the output of time-continuous Gaussian channel with small SNR

Let the process $\xi$ be observed from the sum $\eta$

$$\eta(t) = w(t) + \sqrt{q} \int_0^t \xi(s)ds, \quad 0 \leq t \leq T, \qquad (1)$$

where $w$ is a standard Brownian motion, $\xi$ is any process with zero mean, independent of W, with

$$E_\xi = \int_0^T \xi^2(t)dt < \infty$$

and q is a real constant. Without loss of generality we assume that the signal energy satisfies $E_\xi = 1$.

Relations that connect the *divergence* between the measure in function space of the channel output process and the measure of the Gaussian process with the same covariance, to both CMMSE MMSE, achievable by optimal estimation of the input given the output are well known (e. g. [1]-[3]).

We denote the non-Gaussianess of the channel output $\eta$ by $D(q) = D(\eta \| \tilde{\eta})$, where $D(\eta \| \tilde{\eta})$ is the divergence between the measure induced by $\eta$ in function space and that of $\tilde{\eta}$, where $\tilde{\eta}$ is a Gaussian process with the same covariance function as that of $\eta$.

Based on relations between the divergence and the minimum mean-square errors we show that $D(q)$ is mainly determined for small q by its second derivative at $q=0$.

*Lemma 1:* As $q \to 0$, for *any process* $\xi$ that fulfils (1)

$$D(q) = \frac{1}{2} D^{(2)}(0) \, q^2 + \mathcal{O}(q^3), \qquad (2)$$

where $D^{(n)}(0) = \left. \frac{d^n}{dq^n} D(q) \right|_{q=0}$.

*Proof*

In order to prove the lemma we recall the following relations between the divergence and the minimum mean-square errors [3]:

$$\text{CMMSE}(\tilde{\xi}) - \text{CMMSE}(\xi) = \frac{2}{q} D(q), \quad (3)$$

$$\text{MMSE}(\tilde{\xi}) - \text{MMSE}(\xi) = 2\frac{d}{dq} D(q). \quad (4)$$

In (3), (4) $\tilde{\xi}$ is a Gaussian process with same covariance function as that of $\xi$.

It is clear that as $q \to 0^+$ the values of $\text{CMMSE}(\tilde{\xi})$, $\text{CMMSE}(\xi)$, $\text{MMSE}(\tilde{\xi})$ and $\text{MMSE}(\xi)$ exceed one. Therefore, from (3), (4)

$$D(0) = \frac{dD(q)}{dq}\bigg|_{q=0} = 0. \quad (5)$$

By Taylor's theorem [4, p. 95] equation (5) implies (2).

### III. CMMSE AND MMSE OF A SIGNAL WHICH IS COMPOSED OF THE SUM OF CW INTERFERERS IN ADDITIVE GAUSSIAN NOISE CHANNEL

Let us consider channel (1) with the following non-Gaussian interference signal

$$\xi(t) = \xi_N(t) = \sum_{i=1}^{N} a_i \sqrt{\frac{2}{TN}} \cos(\omega_{k_i} t + \theta_i), \ 0 \le t \le T. \quad (6)$$

In (6) all $a_i$ are mutually independent random variables, independent of w, identically distributed with $Ea_i=0$, $Ea_i^2 = 1$. All phases $\theta_i$ are uniformly distributed, mutually independent and independent of all $a_i$ and w. $\omega_{k_i} = \frac{2\pi k_i}{T}$, $i_1^N$, are circular frequencies. In this case the energy of $\xi$ fulfils for each N $E_\xi = 1$.

Then, the following asymptotic mean square errors for large N hold:

*Proposition 1:* For each fixed q, as $N \to \infty$

$$\text{CMMSE}(\xi_N) = 1 - [\frac{1}{4} + D^{(2)}(0)]\frac{q}{N} + \mathcal{O}(\frac{1}{N^2}), \quad (7)$$

$$\text{MMSE}(\xi_N) = 1 - [\frac{1}{2} + 2D^{(2)}(0)]\frac{q}{N} + \mathcal{O}(\frac{1}{N^2}). \quad (8)$$

In (7), (8) $D^{(2)}(0)$ is as in (2), for the signal (6) where N=1.

The meaning of (7), (8) is that even for unlimited (but fixed) q, both CMMSE and MMSE tend to the signal energy as N goes to infinity with a rate that is proportional to 1/N.

*Proof*

Let us replace the signal $\xi$ in (1) with the signal $\xi_N$ (6), where $a_i = 1, 1 \le i \le N$. Then the following results for the estimation errors $\text{CMMSE}(\xi_N)$ and $\text{MMSE}(\xi_N)$ were proved in [3]:

$$\text{CMMSE}(\xi_N) = \frac{2N}{q} \ln(1 + \frac{1}{2}\frac{q}{N}) - \frac{2}{q} D_N(\eta\|\tilde{\eta}), \quad (9)$$

$$\text{MMSE}(\xi_N) = \frac{1}{1+\frac{q}{2N}} - 2\frac{d}{dq}(D_N(\eta\|\tilde{\eta})). \quad (10)$$

In (9), (10) the divergence $D_N(\eta\|\tilde{\eta})$, is the sum of individual divergences (which are the same for each i, see also [3, section III]). Therefore

$$D_N(\eta\|\tilde{\eta}) = ND(\frac{q}{N}), \quad (11)$$

where $D(\bullet)$ is the diminution of the divergence $D_N(\eta\|\tilde{\eta})$ to the case N=1. The generalization of (9)-(11) to the case of signal with mutually independent, identical distributed random amplitudes $\{a_i\}$ is straightforward.

By (11) and (2), as $N \to \infty$

$$\begin{aligned} D_N(\eta\|\tilde{\eta}) &= ND(\frac{q}{N}) = \frac{N}{2}D^{(2)}(0)(\frac{q}{N})^2 + N\mathcal{O}(\frac{1}{N^3}) \\ &= \frac{1}{2}D^{(2)}(0)\frac{q^2}{N} + \mathcal{O}(\frac{1}{N^2}), \quad N \to \infty. \end{aligned} \quad (12)$$

From (9), (10) and (12) we get (7), (8).

Consider now the following signal that is composed from the sum of *Gaussian* narrowband waves

$$g_N(t) = \sum_{i=1}^{N} \sqrt{\frac{1}{T}}(a_{ci}\cos\omega_{k_i}t + a_{si}\sin\omega_{k_i}t), \quad (13)$$

where $\{a_{ci}, a_{si}\}$, $i_1^N$ are mutually independent, zero mean Gaussian random variables with variance 1/N, and $\omega_{k_i}$, $i_1^N$ are as in (6).

Since $\xi_N$ and $g_N$ possess same covariance we can consider $g_N$ as a realization of $\tilde{\xi}_N$. In order to evaluate the mean-square errors of $g_N$ we can use the fact that both CMMSE and MMSE seek their maximum for the Gaussian signal. Hence, from (9), (10)

$$\text{CMMSE}(g_N) = \frac{2N}{q}\ln(1 + \frac{q}{2N}), \quad (14)$$

$$\text{MMSE}(g_N) = \frac{1}{1+\frac{q}{2N}}. \quad (15)$$

As we can see the estimation errors of the above Gaussian interference also converge to the interference energy with rate that is proportional to 1/N.

Note that the inability to estimate the signal of the above examples for infinite N follows from two consequences. First,

by (12) the measure of the normalized sum of the non-Gaussian narrowband waves becomes "more Gaussian" (in the sense of divergence) as N goes to infinity. Hence, the negative parts in (9), (10) vanish. Second, Gaussian processes are impossible to be estimated if all their spectral components are mutually independent and deeply immersed in the noise. In other words, since all spectral components of the signal are independent (such that there is no mutual information between any disjoint sets of components), they have to be estimated separately. However, in this case each individual snr is low although the total snr is high.

## IV. Non-Gaussianess and MMSE For Low snr

In this section we use a known result for MMSE in order to show non-Gaussianess behavior for small snr.
Let

$$Y = W + \sqrt{q}\, X \qquad (16)$$

Where $X$ is a random variable with $EX^2 < \infty$, independent of $W = \mathcal{N}(0,1)$ and $q$ expresses snr. Without loss of generality we assume $EX = 0, EX^2 = 1$. The Taylor series expansion of the MMSE around $q = 0^+$ to the third order has been obtained in [1] as

$$\mathrm{MMSE}(X,q) = 1 - q + q^2 - \frac{1}{6}[(EX^4)^2 - 6EX^4 \\ - 2(EX^3)^2 + 15]q^3 + \mathcal{O}(q^4) \qquad (17)$$

For more properties of MMSE in the case (16) see [5]. Using lemma 1 we can express $\mathrm{MMSE}(X,q)$ as a Taylor series that is dependent on the derivatives of the non-Gaussianess of $Y$, for small snr. Comparing the resulting expression to (17) above will allow us to find derivatives of $D(q)$ at $q = 0$ as a function of the moments of $X$.

First, let us note that the special case of (1), where

$$\xi(t) = \sqrt{2} X \cos t, \quad 0 \le t \le 2\pi$$

is equivalent to (16) for same $X$. This could be seen immediately by representing signals in function space with the Fourier family basis (one dimensional case). Therefore MMSE and divergence calculations, as well as lemma 1 bring same results in both spaces. Expanding lemma 1 to forth order yields

$$D(q) = \frac{1}{2} D^{(2)}(0) q^2 + \frac{1}{6} D^{(3)}(0) q^3 + \\ \frac{1}{24} D^{(4)}(0) q^4 + \mathcal{O}(q^5) \quad q \to 0 \qquad (18)$$

From (4), (18) and well known result for MMSE of Gaussian random variable where $X$ is replaced in (16) by $\tilde{X} = \mathcal{N}(0,1)$, we get

$$\mathrm{MMSE}(X,q) = \frac{1}{1+q} - [2D^{(2)}(0)q + D^{(3)}(0)q^2 + \\ \frac{1}{3} D^{(4)}(0) q^3 + \mathcal{O}(q^4)] = 1 - [1 + 2D^{(2)}(0)]q + \\ [1 - D^{(3)}(0)]q^2 - [1 + \frac{1}{3} D^{(4)}(0)]q^3 + \mathcal{O}(q^4) \qquad (19)$$

Comparing (19) with (17) leads to the following proposition

*Proposition 2:* The non-Gaussianess $D(q)$ of the sum (16) of any random variable with zero mean and unit variance, and a Gaussian variable $\mathcal{N}(0,1)$, satisfies:

$$D(0) = D^{(1)}(0) = D^{(2)}(0) = D^{(3)}(0) = 0 \qquad (20)$$

$$D^{(4)}(0) = \frac{1}{2}[(EX^4)^2 - 6EX^4 - 2(EX^3)^2 + 9] \qquad (21)$$

Hence, in the special case of random variables (16), no matter how large is the non-Gaussianess of $X$, the first four terms in the Taylor expansion of the non-Gaussianess of the sum $Y$ equal zero.